\documentclass[amsmath,amssymb,superscriptaddress,nobalancelastpage,aps,longbibliography,twocolumn,showpacs]{revtex4-1}

\usepackage{varioref}
\usepackage{xr-hyper}
\usepackage{xcolor}
\usepackage{nicefrac}
\usepackage{xfrac}
\usepackage{hyperref}
\hypersetup{colorlinks,linkcolor=blue,urlcolor=blue,citecolor=orange}
\usepackage{ulem}
\usepackage{siunitx}
\usepackage{graphicx}% Include figure files
\usepackage{dcolumn}% Align table columns on decimal point
\usepackage{bm}% bold math

\DeclareMathOperator\erf{erf}

\newcommand{\ybcosss}{$\rm YBa_2Cu_3O_{6.67}$}

\usepackage{color}  % remove for submission

\newcommand{\cut}[1] {\textcolor{green}{[----]}}

\begin{document}

\title{Evidence for fragile superconductivity in the presence of disordered charge density waves in YBa$_{2}$Cu$_{3}$O$_{6.67}$}

\title{Phase-Competition-Induced Two-Flavor Superconductivity in YBa$_2$Cu$_3$O$_{6.67}$}

\title{Momentum Differentiated Phase Competition in YBa$_2$Cu$_3$O$_{6.67}$}

\title{Spatially Inhomogeneous Competition between Superconductivity and the Charge Density Wave in YBa$_2$Cu$_3$O$_{6.67}$}

\author{J. Choi}
\affiliation{Physik-Institut, Universit\"{a}t Z\"{u}rich, Winterthurerstrasse 190, CH-8057 Z\"{u}rich, Switzerland}

\author{O. Ivashko}
\affiliation{Physik-Institut, Universit\"{a}t Z\"{u}rich, Winterthurerstrasse 190, CH-8057 Z\"{u}rich, Switzerland}
\affiliation{Deutsches Elektronen-Synchrotron DESY, 22607 Hamburg, Germany.}

 \author{E. Blackburn}
 \affiliation{School of Physics and Astronomy, University of Birmingham, Birmingham B15 2TT, United Kingdom.}
\affiliation{Division of Synchrotron Radiation Research, Department of Physics, Lund University, S\"{o}lvegatan 14, 22100 Lund, Sweden}

\author{R. Liang}
\affiliation{Department of Physics $\&$ Astronomy, University of British Columbia, Vancouver, Canada.}
\affiliation{Canadian Institute for Advanced Research, Toronto, Canada.}

\author{D. A. Bonn}
\affiliation{Department of Physics $\&$ Astronomy, University of British Columbia, Vancouver, Canada.}
\affiliation{Canadian Institute for Advanced Research, Toronto, Canada.}

\author{W. N. Hardy}
\affiliation{Department of Physics $\&$ Astronomy, University of British Columbia, Vancouver, Canada.}
\affiliation{Canadian Institute for Advanced Research, Toronto, Canada.}

\author{A.~T. Holmes}
\affiliation{European Spallation Source ERIC, Box 176, SE-221 00 Lund, Sweden}

\author{N.~B. Christensen}
\affiliation{Department of Physics, Technical University of Denmark, DK-2800 Kongens Lyngby, Denmark.}

\author{M. H\"ucker}
\affiliation{Department of Condensed Matter Physics, Weizmann Institute of Science, Rehovot 7610001, Israel}
%\affiliation{Condensed Matter Physics \& Materials Science Dept., Brookhaven National Lab., Upton, NY 11973, USA.}

\author{S. Gerber}
\affiliation{Laboratory for Micro and Nanotechnology, Paul Scherrer Institut, Forschungsstrasse 111, CH-5232 Villigen PSI, Switzerland}

\author{O. Gutowski}
\affiliation{Deutsches Elektronen-Synchrotron DESY, 22607 Hamburg, Germany.}

\author{U. R\"utt}
\affiliation{Deutsches Elektronen-Synchrotron DESY, 22607 Hamburg, Germany.}

 \author{M.~v.~Zimmermann}
\affiliation{Deutsches Elektronen-Synchrotron DESY, 22607 Hamburg, Germany.}

 \author{E. M. Forgan}
 \affiliation{School of Physics and Astronomy, University of Birmingham, Birmingham B15 2TT, United Kingdom.}

 \author{S. M. Hayden}
 \email{S.Hayden@bristol.ac.uk}
\affiliation{H. H. Wills Physics Laboratory, University of Bristol, Bristol, BS8 1TL, United Kingdom.}

\author{J. Chang}
\email{johan.chang@physik.uzh.ch}
\affiliation{Physik-Institut, Universit\"{a}t Z\"{u}rich, Winterthurerstrasse 190, CH-8057 Z\"{u}rich, Switzerland}

\date{\today}

\begin{abstract}
The charge density wave in the high-temperature superconductor YBa$_2$Cu$_3$O$_{7-x}$ (YBCO) has two different ordering tendencies differentiated by their $c$-axis correlations. These correspond to ferro- (F-CDW) and antiferro- (AF-CDW) couplings between CDWs in neighbouring CuO$_2$ bilayers. This discovery has prompted several fundamental questions: how does superconductivity adjust to two competing orders and are either of these orders responsible for the electronic reconstruction? Here we use x-ray diffraction to study YBa$_2$Cu$_3$O$_{6.67}$ as a function of magnetic field and temperature. We show that regions with F-CDW correlations suppress superconductivity more strongly than those with AF-CDW correlations. This implies that an inhomogeneous superconducting state exists, in which some regions show a fragile form of superconductivity. By comparison of F-CDW and AF-CDW correlation lengths, it is concluded that F-CDW ordering is sufficiently long-range to modify the electronic structure. Our study thus suggests that F-CDW correlations impact both superconducting and normal state properties of YBCO.

\end{abstract}

\maketitle

\textit{Introduction:} Many theories and experiments point to an inhomogeneous nature of the superconductivity of high-temperature cuprate superconductors (HTC). Indeed, inhomogeneity in the sense of phase seperation, intertwining of competing order parameters, stripes or pair-density-waves \cite{Zaanen1989,Kivelson03a,BergPRL2007,Machida89,Emery93} may be at the heart of cuprate  superconductivity. Underdoped high-temperature superconductors exhibit competing tendencies towards charge-density-wave (CDW) and superconducting (SC) orders~\cite{Tranquada95a,Wu11a,Ghiringhelli12a,Chang12a,LeBoeuf13a}. This means that the effects of quenched disorder may induce defects that disrupt or suppress the CDW order, and  in turn lead to a resurgence of superconductivity.
%in turn resurgence superconductivity. 
The application of a magnetic field provides another interesting control parameter because the introduction of vortices into the superconducting state tips the energy balance in favour of CDW order.  

\begin{figure*}
\center{\includegraphics[width=\textwidth]{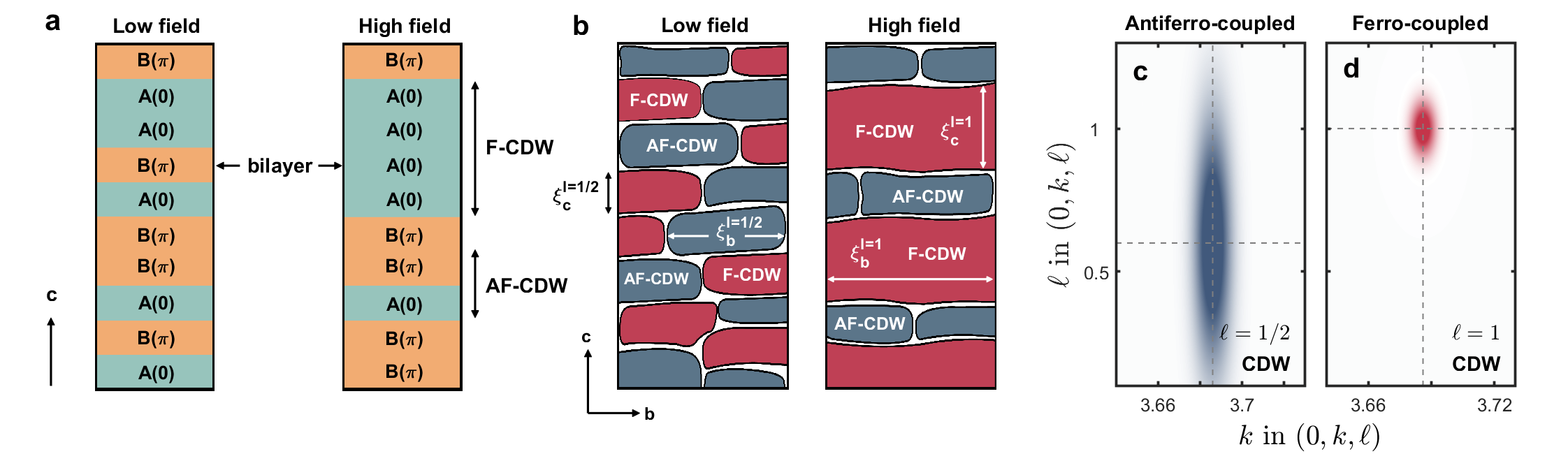}} \caption{\textbf{Schematic real-space illustration of the disordered charge density wave (CDW) in YBa$_2$Cu$_3$O$_{6.67}$.} (\textbf{a}) Representative CDW stacking sequences along the $c$-axis at low and high magnetic field for a fixed point in the $a$-$b$ plane. A and B represent the two possible phases in the CDW modulation in each bilayer \cite{Chang16}. (\textbf{b}) Extension to the $b$-$c$ plane. Spatially separated regions where the F-CDW ($\ell=1$) and AF-CDW ($\ell=1/2$) correlations in (a) are present. (\textbf{c,d}) Idealized diffraction intensity for the F-CDW and AF-CDW correlations at high field.}
\label{fig:realspace}
\end{figure*}

In this context, YBa$_2$Cu$_3$O$_{6+x}$ (YBCO) is an important model system where large onset temperatures of both superconductivity and CDW order are present~\cite{Ghiringhelli12a,Chang12a,Achkar12a,wu15}. As a function of both magnetic field and uniaxial stress, two %distinct 
intimately related CDW ordering tendencies have been realized~\cite{Gerber15,Chang16,Jang16,KimScience2018}. These correspond to different ordering patterns along the $c$-axis~\cite{Gerber15,Chang16}, as schematically illustrated in Fig. \ref{fig:realspace}. Magnetic field (or uniaxial stress) induces an in-phase ferro-coupled CDW (F-CDW) along the $c$-axis~\cite{Chang16,Jang16} on top of the original out-of-phase bi-axial antiferro-coupled CDW (AF-CDW) order. Quenched disorder, that in YBCO is naturally introduced through imperfections in the chain layer, may have important implications for the CDW order~\cite{Jang16,YuPRB2019,CaplanPRL2017}. It is believed that disorder locally favors the AF-CDW order and that the evolution of the CDW correlations with magnetic field may be understood as a crossover transition in which the CDW coupling along the $c$-axis varies~\cite{Chang16,CaplanPRL2017}. The introduction of normal vortices with field leads to a more ordered CDW. This suggests disorders associated with vortices is of secondary importance.

More recently, the case where competing tendencies towards CDW and SC order in the presence of topological defects due to quenched disorder and of vortices in the SC state has been considered~\cite{YuPRB2019}. It was demonstrated theoretically, that a fragile SC state can occur at low temperatures and at high magnetic field.  This state is based on regions in which the CDW order is weakened due to defects where locally superconducting halos can form. These regions then can couple to form a state with global SC phase coherence. Given the richness of theoretical possibilities  when superconductivity competes with CDW order in presence of vortices and weak disorder, it is of great interest to scrutinize the magnetic field-induced phase competition in YBCO.

Here we present a comprehensive study of the F-CDW and AF-CDW correlations in YBa$_2$Cu$_3$O$_{6.67}$. By varying the magnetic field strength $B$ -- applied along the crystallographic $c$-axis -- and thermal excitation energy $T$, new insights into in-plane correlation lengths in the phase space $(B,T)$ and competition of the correlations with superconductivity are obtained. F-CDW correlations are traced from low fields. From the diffraction intensities of F- and AF-CDW correlations versus temperature and magnetic field, the competition with superconductivity is studied. In particular, we extract the characteristic temperatures below which the respective CDW correlations are suppressed. % are extracted. 
%These temperatures indicate the onset of superconductivity in regions of the sample with F-CDW and AF-CDW correlations. They are significantly different for high fields where the F-CDW correlation length becomes large.
They are significantly different at high magnetic fields where the F-CDW correlation length becomes large and the F-CDW correlations are suppressed at a significantly lower temperature scale. These different temperatures indicate where superconducting order becomes strong enough to influence the F-CDW and AF-CDW correlations. This observation suggests that superconductivity in the F-CDW regions is more strongly suppressed by the CDW. In this fashion, inhomogeneous superconductivity emerges due to competition with CDW correlations which are locally different. Consequently, regions with weaker or more fragile superconductivity are created. Another implication of our study is that in-plane correlation lengths can be deduced across the $(B,T)$ phase space. 
%\blue{Although not identical,} 
We extract the in-plane correlation lengths at the field-dependent temperature scales set by the phase competition between superconductivity and the two CDW ordering tendencies. It is demonstrated that the F-CDW in-plane correlation length exceeds that of the AF-CDW order down to magnetic fields as small as 5~T. 
%This result implies that F-CDW correlations may be responsible for the observed Fermi surface reconstruction~\cite{LeBoeuf07a}.
This result raises the question as to whether F-CDW correlations may be related in the Fermi surface reconstruction. Together with the recent finding that uniaxial strain induces F-CDW order~\cite{KimScience2018}, our results suggest that F-CDW order in YBCO plays an important role for both the electronic structure and in rendering superconductivity into an inhomogeneous state that includes a more fragile flavor. \\[1mm]

\begin{figure*}
\center{\includegraphics[width=\textwidth]{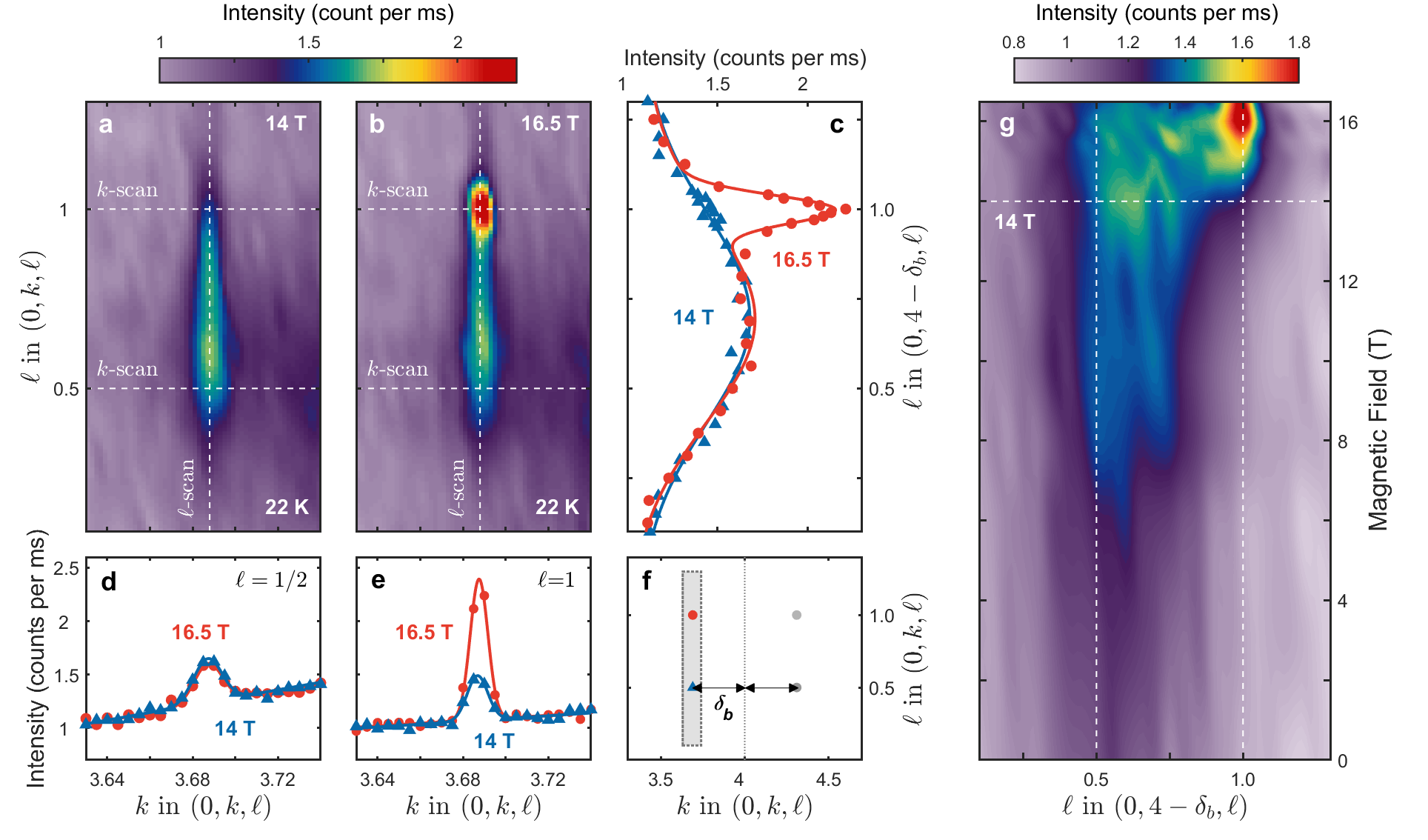}} \caption{\textbf{Magnetic-field-induced correlations and three-dimensional CDW order in YBa$_2$Cu$_3$O$_{6.67}$.} (\textbf{a,b}) X-ray diffraction intensity maps around $(0,4-{\delta}_b, \ell)$ with  $B=14$~T and  $16.5$~T, respectively.
%(b) reveal incommensurate lattice modulations at $T=22$ K. 
(\textbf{c,d}), and (\textbf{e}) are $\ell$- and $k$-dependent intensity profiles along the vertical and horizontal white dashed lines in (\textbf{a,b}).
%(d) and (h) display %(background) subtraction of the scattering observed at 14 T. 
(\textbf{f}) delineation of the scattering plane and probed reciprocal space (grey). The in-plane incommensurability $\delta_b$ is defined by the horizontal arrow. (\textbf{g}) diffraction intensity map displayed in false color, as a function of $(4,0,\ell)$ and magnetic field. Source data are provided as a Source Data file.}
\label{fig:rods}
\end{figure*}

\begin{figure*}
\center{\includegraphics[width=0.7\textwidth]{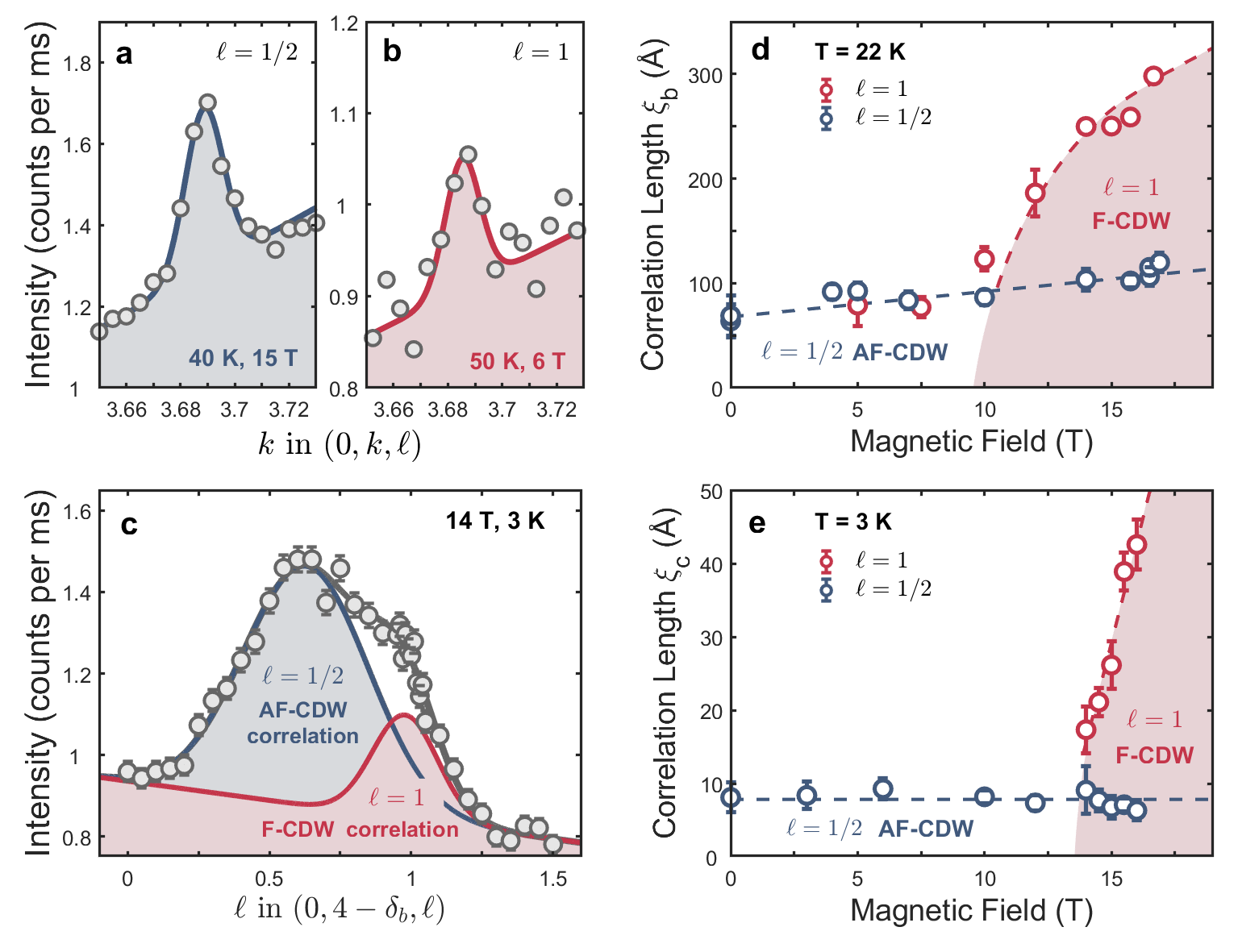}}\caption{\textbf{Two-component analysis of out-of-plane charge-density-wave (CDW) correlations.} %$(0,4-{\delta}_{b},1)$. 
%(\textbf{a}) Schematic real-space illustration of spatially separated regions where the F-CDW ($\ell=1$) and AF-CDW ($\ell=1/2$) correlations are present. (\textbf{b,c}) Idealized diffraction intensity for the F-CDW and AF-CDW correlations. 
%in a reciprocal space.
(\textbf{a,b}) Raw diffraction intensities along $(0,k,1/2)$ and $(0,k,1)$ for temperatures and magnetic fields as indicated.
%with $\ell=0.5$ and 1
(\textbf{c}) Raw diffraction intensity profile along the $(0,4-\delta_b,\ell)$ direction, fitted with two Gaussian functions on a linear background. The dark grey line is their sum. (\textbf{d,e}) Magnetic-field-dependent evolution of the respective correlation lengths for $\ell=1/2$ and 1. The dashed lines and shaded area are guides to the eye. Error bars are standard deviations determined by counting statistics and Gaussian fits, respectively. Source data are provided as a Source Data file.
%The grey- and red-colored area indicate the region for a crossover and 3D CDW order, respectively. 
}\label{fig:two_component}
\end{figure*}
 
\textbf{Results}\\
\textbf{Charge-density-waves in YBCO.}
Charge density wave (CDW) ordering has been found to be a universal property of hole doped cuprates~\cite{Tranquada95a,Wu11a,Ghiringhelli12a,Chang12a,Comin14,daSilvaNeto14,Tabis14}. YBCO has a bilayer structure with relatively close ($d= 3.3$~\AA) pairs of CuO$_2$ planes, %known as bilayers 
separated by layers containing CuO chains running along the $b$-axis.
In YBCO, the CDW results in weak scattering centred at $\textbf{Q}_{\textrm{CDW}}(\ell)=\boldsymbol{\tau} + \textbf{q}_{\textrm{CDW}} + \ell \textbf{c}^{\star}$, where $\boldsymbol{\tau}$ is a reciprocal lattice point of the unmodulated structure and $\textbf{q}_{\textrm{CDW}}=(\delta_a,0,0)$ or $(0,\delta_b,0)$~\cite{Ghiringhelli12a,Chang12a,Achkar12a}.  For the YBa$_2$Cu$_3$O$_{6.67}$ ($T_c$ = 67~K) composition studied here, two charge-density-wave components have been identified~\cite{Gerber15,Chang16,Jang16}. 
The first found in zero magnetic field develops below $T_{\textrm{CDW}}=140(10)$~K with $\delta_a=0.305(2)$ and $\delta_b=0.314(2)$. 
The phase of this CDW in neighbouring bilayers has an anti-parallel correlation with only weak correlation along the $c$-axis, manifested by a broad intensity peak with an $\ell\approx 1/2$ modulation. 
%it is found \cite{Ghiringhelli12a,Chang12a} that the modulated rods of scattering develop below $T_{\textrm{CDW}}=140(10)$~K with $\delta_a=0.305(2)$ and $\delta_b=0.314(2)$. 
The intensity of these rods of scattering is enhanced by the application of a magnetic field along the $c$-axis~\cite{Chang12a,Blackburn13a,Huecker14}. Fig.~\ref{fig:rods}a displays this rod in the $(0,k,\ell)$ scattering plane for $B=14$~T.
A second magnetic-field-induced CDW exists along the $b$-axis direction~\cite{Gerber15,Chang16,Jang16}. This CDW has the same phase across neighbouring bilayers (ferro-coupled) and hence is manifested by a $\ell=1$ reflection on top of the $\ell \sim 1/2$ scattering rod (see Fig.~\ref{fig:realspace}c,d and Fig.~\ref{fig:rods}a,b). In this paper we use the term ``CDW order" loosely to mean static CDW correlations with a finite correlation length which develop at low temperatures and high magnetic fields.

%YBCO has a bilayer structure with relatively close ($d= 3.3$~\AA) pairs of CuO$_2$ planes known as bilayers separated by planes containing CuO chains.  
%For $B=0$, the phase of the CDW in neighbouring bilayers is anti-correlated, that is, there is a phase change of $\pi$ and the correlation length is $\eta_c^{AC} \approx 20$~\AA. The short correlation length along the along the $c$-axis has led to the zero field CDW being described as ``two-dimensional'' (2D). A better description is to say that the CDW is anti-correlated between bilayers or has $\ell=1/2$-order.  The CDW in zero magnetic field is bi-directional in that it has propagation vectors $\boldsymbol{\delta}_a$ and $\boldsymbol{\delta}_b$ along the $\mathbf{a}$ and $\mathbf{b}$ axes, where the CuO chains are along $\mathbf{b}$. 
%The application of a magnetic field along the $\mathbf{b}$ axis increases the amplitude of both wavevector components ($\mathbf{q}_a$ and $\mathbf{q}_b$) of the CDW.  In addition, the nature of $\mathbf{q}_b$ correlations changes along the $c$-axis, with regions of the sample changing such that the phase of the CDW is the same in neighbouring bilayers i.e. they exhibit $\ell=1$-order.  As the magnetic field is increased the size of the $\ell=1$-ordered regions increases. 

Modelling \cite{Chang16} of the $c$-axis correlations in YBCO suggests that magnetic field changes the inter-bilayer (IB) coupling of the CDW. This is an indirect effect in that the field suppresses the superconductivity which changes the IB coupling. The CDW will also be pinned by defects in the chain layer. Thus, we can view the evolution of the CDW with field as a crossover transition. At low field $B \lesssim 14$~T, the CDW is pinned and the system displays weakly correlated $\ell=1/2$ order. At very high field ($B \gg 14$~T), 
%the coupling along the $c$-axis favours in-phase coupling of the bilayers, this coupling is sufficient to overcome the pinning and forms a state with long range $l=1$ order along the $b$-axis only.  \\[2mm]
in-phase coupling of the bilayers is favoured, which is sufficient to overcome the pinning and forms a state with long range $l=1$ order along the $b$-axis only.  \\[2mm]

\textbf{Two-component analysis.} 
The two CDW orderings can be analysed separately as a function of temperature and magnetic field. Here we employ three separately independent but consistent methodologies for data analysis. First, as shown in Figs.~\ref{fig:rods}d,e and Figs.~\ref{fig:two_component}d,e, the in-plane correlations and diffraction intensities can be inferred from Gaussian fits versus $\mathbf{Q}_b$ at $\ell=1/2$ and $\ell=1$. The correlation lengths are defined as $\xi^{\ell=1/2,1}_b=1/\sigma^{\ell=1/2,1}_b$ where $\sigma^{\ell=1/2}_b$ and $\sigma^{\ell=1}_b$ are the respective standard deviations. Second, along the $(0,4-\delta_b,\ell)$ direction, the out-of-plane correlation lengths $\xi^{\ell=1/2,1}_c$ can be deduced by fitting with a double Gaussian function (Fig.~\ref{fig:two_component}c). 
%Although the fitted peak positions are found to be $\ell\approx0.6$ and 1, we label the correlation length as $\xi^{\ell=1/2,1}_c$.  
The centre in $\ell$ of the broad AF-CDW signal rises slightly with magnetic field \cite{Chang16} (the values that were fitted and at high field were all $\sim$0.6), however, for simplicity we label the correlation lengths as $\xi^{\ell=1/2,1}_c$. We find that the temperature dependence of the broad peak is insensitive to the exact $\ell$ value (see supplementary Fig.~1), and all $h$- and $k$-scans were carried out at $\ell=1/2$.
%Based on the detailed intensity profile along $\ell$, the two Gaussian peak positions were set to $\ell=0.6$ and 1, respectively. 
Third, as the $\ell\approx1/2$ CDW component has essentially no magnetic field dependence in the narrow range 14-16 T (Fig.~\ref{fig:rods}c), it is possible -- by subtraction of Gaussian fits of 14~T diffraction intensities -- to isolate the field and temperature dependence of the $\ell=1$ diffraction intensity. \\

\begin{figure*}
\center{\includegraphics[width=0.9\textwidth]{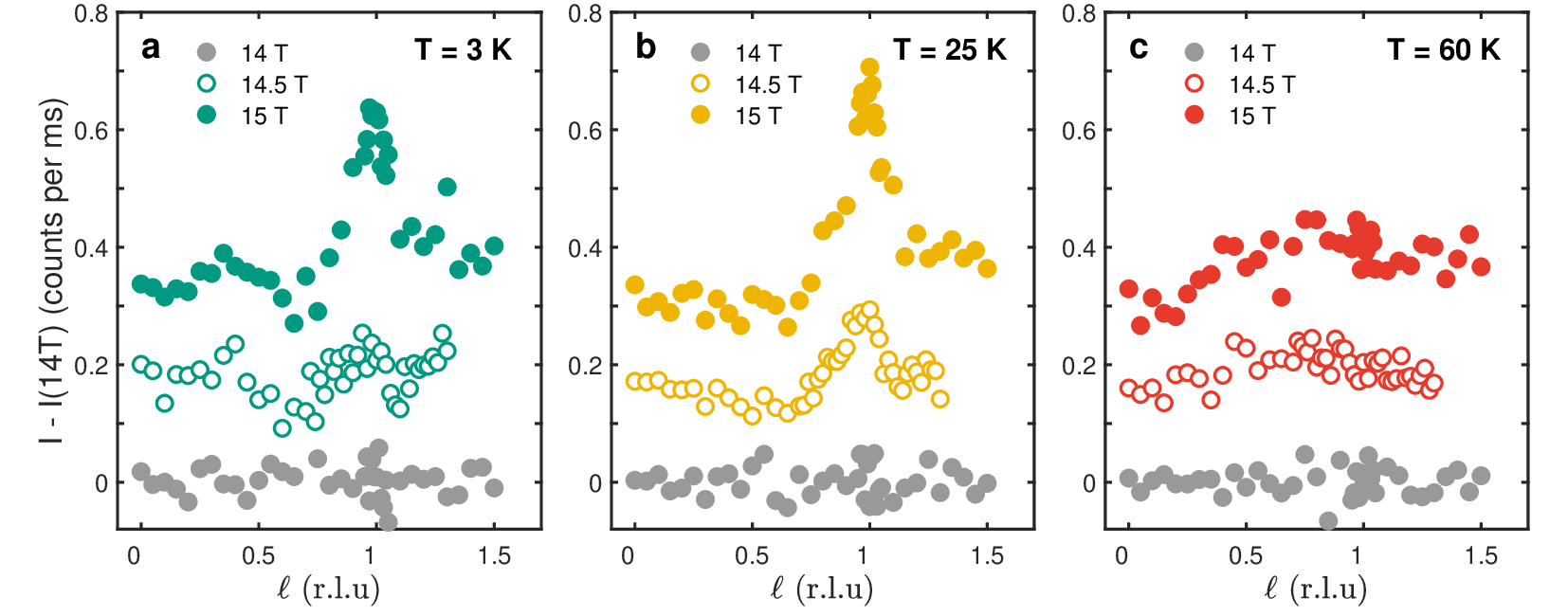}}\caption{\textbf{Temperature and field dependence 
of in- and out-of-plane diffraction intensities.} \textbf{(a-c)} $\ell$-dependent diffraction intensity after subtracting the Gaussian fitted $B=14$~T profiles at temperatures and magnetic fields as indicated. In this fashion, the relative field effects above 14 T are extracted. Raw 14~T $\ell$-scans are shown in supplementary Fig.~2. Data at different fields have been given an arbitrary shift for the sake of presentation. Source data are provided as a Source Data file.
%(\textbf{d,e}) Peak amplitude $I$ extracted from $k$-scans through $(0,4-{\delta}_{b},\ell)$ with $\ell=0.5$ and 1, and plotted versus temperature at various magnetic fields
%as indicated. (\textbf{f}) Same data as in (\textbf{e}) but with an arbitrary vertical shift for clarity.
%Arrows in (\textbf{e,f}) indicate the temperature scale $T_{max}^{3D}$ below which suppression of CDW order is observed. 
%Error bars are s.d.’s of the fit parameters described in the text.
%The data is vertically shifted in (c) for visual clarity.
%(c) Background subtracted $I-I(14T)$ response at $\ell=1$ displayed as a function of temperature for $H>14$ T.
% approximate onset temperature of $\ell=1$ charge ordering is indicated by the gray-shaded band. 
}\label{fig:intensities}
\end{figure*}

\begin{figure*}
 	\begin{center}
 		\includegraphics[width=0.9\textwidth]{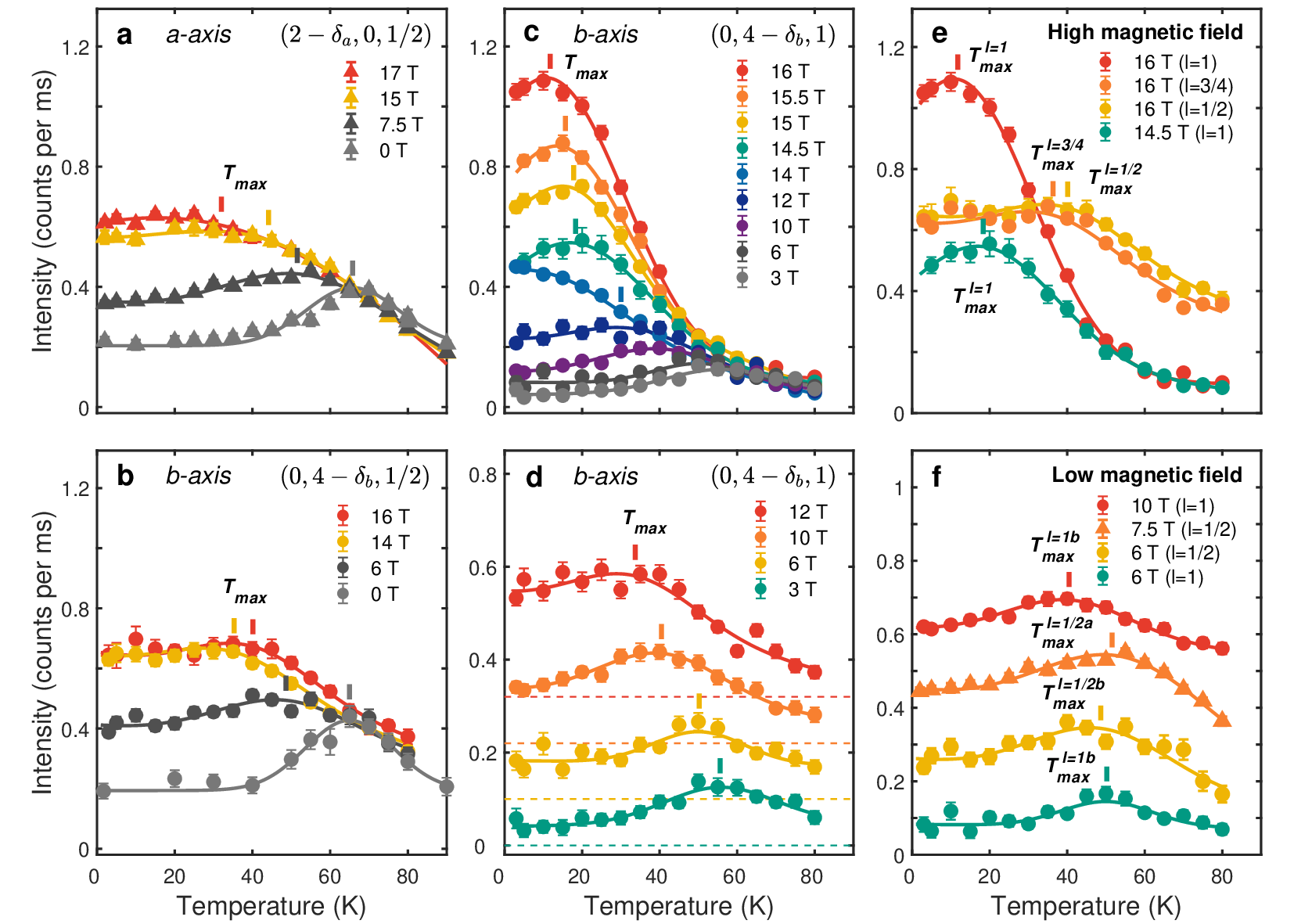}
 	\end{center}
 	\caption{\textbf{Temperature scales for phase competition:} \textbf{(a-f)} Peak intensities versus temperature measured at $\ell=1/2$ and 1 charge-density-wave modulation vectors versus temperature. \textbf{(a,b)} Intensity for $(2-\delta_a,0,1/2)$ and $(0,4-\delta_a,1/2)$ with magnetic fields as indicated. Data in (a) and zero-field data in (b) are re-plotted from Ref.~\onlinecite{Chang12a}. Non-zero magnetic field data in (b) is $\ell\sim 0.6$ amplitudes obtained from two-Gaussian fits of $\ell$-scans. \textbf{(c)} Diffraction intensities at $(0,4-\delta_a,1)$. \textbf{(d)} same data as in (c) but zoomed on the low-field curves. For visibility, these curves have been given an arbitrary vertical shift -- as indicated by the dashed base-lines. \textbf{(e,f)} comparisons of $\ell=1$ and $1/2$ temperature dependencies and with that the associated maximum temperature scales $T_{\textrm{max}}$ for the high- and low- magnetic field regimes respectively. Solid lines are fits to a skewed Gaussian (see text) on a linear background. Vertical bars are the $T_\textrm{max}$ temperature scales defined by the maximum of the second derivative of the skewed Gaussian. Error bars are standard deviations due to counting statistics. Source data are provided as a Source Data file.}\label{fig4}
\end{figure*}

\textbf{Superconductivity and CDW phase competition.} 
By subtracting the Gaussian fit of 14-T diffraction intensity from those measured at high fields (see supplimentary Figs.~2 and 3), the temperature dependence of the $\ell=1$ CDW component at high field is deduced. For both $B=14.5$ and 15~T, the diffraction intensity is maximum at a finite temperature of $T\approx 25$~K (see Fig.~\ref{fig:intensities}). It thus demonstrates that the CDW intensity at $\ell=1$ has a non-monotonic temperature dependence. This effect can also be analysed from in-plane intensity scans.
In Fig.~\ref{fig4}a-f, we plot the peak intensity $I$ of Gaussian fits to $k$-scans through $(0,4-\delta_b,\ell)$ as a function of temperature and at magnetic fields as indicated. At $\ell=1/2$ (Fig.~\ref{fig4}), the zero-field temperature dependence displays a maximum at $T=T_{\textrm{max}}^{\ell=1/2}=T_\textrm{c}$ -- consistent with previous reports~\cite{Chang12a,Ghiringhelli12a}. 
%\pink{As the magnetic field is increased, $T_{\textrm{max}}^{\ell=1/2}(B)$ moves to lower temperature along the $a$-axis direction (Fig.~\ref{fig4}a). However, this effect is less evident along the $b-$axis indicating possible anisotropy in $T_{max}$ along two directions (Fig.~\ref{fig4}b).}
As the magnetic field is increased, $T_{\textrm{max}}^{\ell=1/2}(B)$ moves to lower temperature. This effect is found along both $a-$ and $b-$ axis directions (Fig.~\ref{fig4}a,b).

\begin{figure*}
\center{\includegraphics[width=1.0\textwidth]{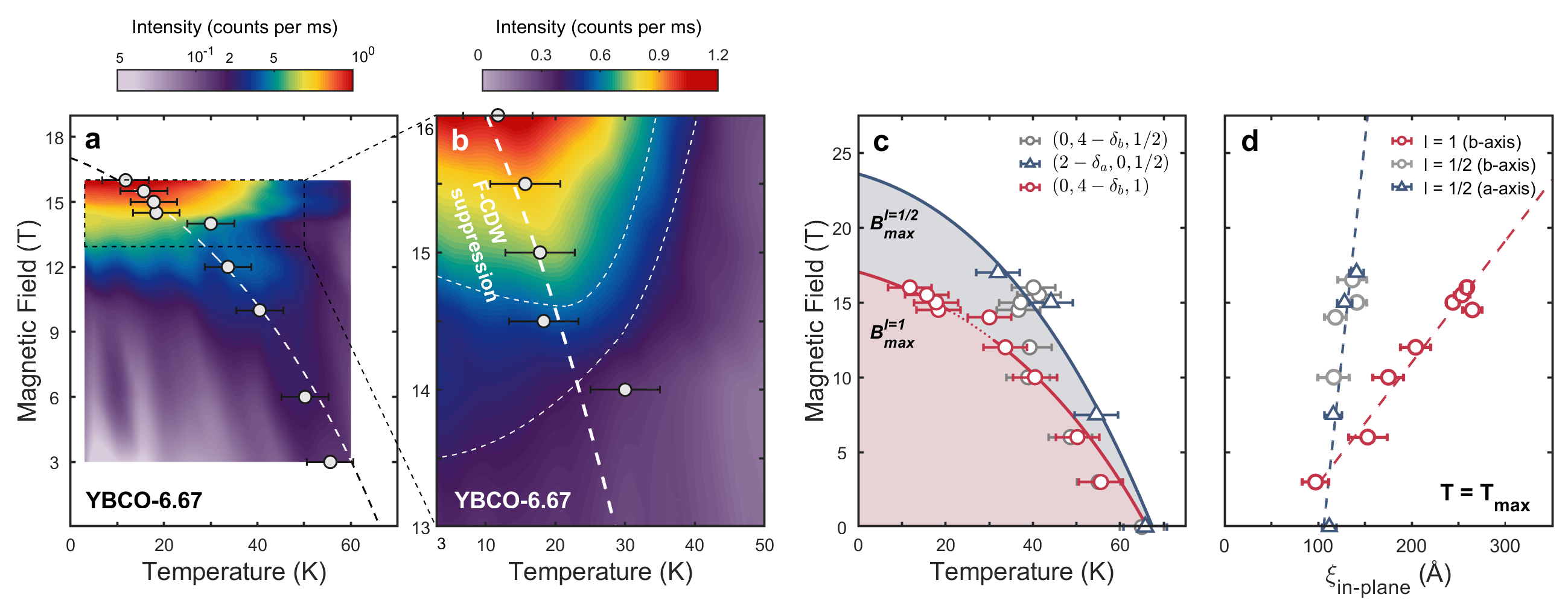}} \caption{\textbf{Competing temperature and magnetic field scales.} \textbf{(a)} Intensity map of (0,$4-\delta_b$,1) (logarithmic color scale) as a function of temperature and magnetic field. The open circles indicate the temperature scale $T_\textrm{max}^{\ell=1}$ below which suppression of the F-CDW order is observed (Fig.~\ref{fig:intensities}a-c). The curved dashed line is a guide to the eye. \textbf{(b)} Intensity map (linear color scale) focusing on the high-magnetic-field region. The thin dashed lines indicate contour lines. \textbf{(c)} $T_\textrm{max}$ scales inferred from the $\ell=1/2$ (red) and $\ell=1$ (blue) charge-density-wave, respectively. Solid lines are guides to the eye. For the $\ell=1/2$ $T_\textrm{max}$ temperature scale, triangular and circular data points stem from data recorded along the $a$- (Ref. \onlinecite{Chang12a}) and $b-$axis (this work) respectively. As the $T_\textrm{max}$ temperature scales indicate the onset of competition with superconductivity, these lines are labeled $B_\textrm{c2}^{\ell=1}$ and $B_{c2}^{\ell=1/2}$ in the $T\rightarrow 0$ limit. 
%For completeness a upper critical field $H_{c2}$ line in absence of charge-density-wave order is indicated schematically by the grey line. 
Error bars indicate the uncertainty related to the determination of the $T_\textrm{max}$ temperature scales. \textbf{(d)} Correlation lengths $\xi_b^{\ell=1/2}$ and $\xi_b^{\ell=1}$ at the respective $T_\textrm{max}$ temperature values. Source data are provided as a Source Data file.
}\label{fig:maps}
\end{figure*}

The behaviour of the $\ell=1$ intensity is very different. For low fields, $B \lesssim 10$~T, the field induces a relatively weak response at $\ell=1$. It shows a maximum at similar temperature as the $\ell=1/2$ component (see Fig.~\ref{fig4}d). For $10 \lesssim B \lesssim 16$~T, there is a rapid  increase of $I(\ell=1)$ at low temperature (Fig.~\ref{fig4}c).  
The rapid temperature dependence below 40~K associated with the crossover field may obscure any maximum due to competition with superconductivity.
%At the ``crossover field'' $B=14$~T,  $I(\ell=1)$ increases monotonically with decreasing temperature, \textit{i.e.} no maximum is observed. 
Then at higher fields $B \ge 14.5$~T, a maximum again develops at $T_{\textrm{max}}^{\ell=1}$ which is considerably lower than $T_{\textrm{max}}^{\ell=1/2}$ for the same field. Direct comparisons of the high- and low-field diffraction intensities at $\ell=1/2$ and $\ell=1$ versus temperature are shown in Fig.~\ref{fig4}e,f. To extract $T_{\textrm{max}}^{\ell=1/2}$ and $T_{\textrm{max}}^{\ell=1}$ in a systematic fashion where the maximum is less well defined, the diffraction intensity versus temperature curves are fitted using a skewed Gaussian function $\Gamma(T)=G(T)[1+\erf(\alpha T/\sqrt{2})]$ on a linear background-- solid lines in Fig.~\ref{fig4}. Here $G(T)$ is a Gaussian function and $\erf(\alpha T)$ is the error function.  The characteristic temperatures, shown in Fig.~\ref{fig:maps}c versus magnetic field, are derived from the criterion $T_\textrm{max}=\textrm{max}(-d^2\Gamma/dT^2)$ --
%that selects either a maximum or an inflection point in the diffraction intensity versus temperature. We view $T_{max}$ as a characteristic temperature scale for competition between CDW and superconducting order.\\
that selects either a maximum or saturation point in the diffraction intensity versus temperature. We view $T_\textrm{max}$ as a characteristic temperature scale for competition between CDW and superconducting order.\\

%Direct comparisons of the high- and low-field scattering amplitudes at $\ell=0.5$ and $\ell=1$ versus temperature are shown in Fig.~4e,f and the $T_{\textrm{max}}$'s are displayed versus magnetic field in Fig.~5c.\\

%This indicates that the regions of the sample contributing the $\ell=1$ component are behaving differently to those contributing to $\ell=1/2$. 

%The $\ell=1$ cusp-like temperature dependence is also encoded into the intensity versus $\ell$ profiles. As shown in Fig. 1 and 3, the $\ell=0.5$ response is essentially field independent in the narrow window 14-16 Tesla. Subtracting the $\ell$-profile at 14~T is therefore another way to isolate the $\ell=1$ response. \\

%As shown in Fig. 3, the $\ell=1$ intensity (after subtraction) is peaking at finite temperatures confirming qualitatively the cusp-like temperature dependence. 

\textbf{Correlation lengths.}
Next, we turn to discuss the CDW correlation lengths. Figs. \ref{fig:two_component}d,e show correlation lengths $\xi_b$ and $\xi_c$ for the lowest temperatures we measured: $T=22$ and 3~K, respectively. The out-of-plane correlation length $\xi^{\ell=1/2}_c\sim 10$~\AA\ is essentially magnetic-field independent  while the in-plane correlation length $\xi^{\ell=1/2}_b\sim 60$~\AA\ increases weakly.
%(see Fig. \ref{fig:two_component}g,h). 
This is in strong contrast to the $\ell=1$ order that gains a dramatic out-of-plane ``coherence'' above 14 T (Fig. \ref{fig:two_component}e). 
In fact, $\xi^{\ell=1}_c$ undergoes a four-fold increase %when going 
from 14 to 16 T. Prior to this change, %the in-plane correlations 
$\xi^{\ell=1}_b$  increases
significantly for field strengths above 10 Tesla (Fig. \ref{fig:two_component}d). 
Thus at high-fields, the correlation length of the $\ell=1$ order exceeds that of the $\ell=1/2$ CDW component in both in- and out-of-plane directions.
It is also interesting to discuss the in-plane correlations at the  
$T_\textrm{max}$ temperature scales.
%that, although different for $\ell=1/2$ and 1, delineates a maximum in the respective diffraction intensities. 
For $\ell=1/2$, the correlation length $\xi^{\ell=1/2}_b$ at $T_\textrm{max}$, is about 20\% longer
than that at $T=22$~K in the studied field range (compare Fig.~\ref{fig:two_component}d and \ref{fig:maps}d).
The $\ell=1$
in-plane correlations extend down to zero-magnetic-field (Fig.~\ref{fig:maps}d).
At $T_\textrm{max}$
%In contrast to $T=22$~K, the 
$\xi_b^{\ell=1}$ 
%in-plane correlations 
exceeds 
$\xi^{\ell=1/2}_b$ %already 
for $B>5$~T (Fig.~\ref{fig:maps}d), while at 22~K this is only seen at 10~T (Fig.~\ref{fig:two_component}d). 
%In summary, consideration of correlation lengths versus temperature and magnetic field suggests that the F-CDW correlations compete more strongly with superconductivity.\\
It is of great interest to note the interplay between the observed peak intensities and the correlation lengths. Our estimates of the k-space integrated intensity for the $\ell = 1$ peak show approximately constant values versus field above 14 T (See Supplementary Figure 4). This implies that the F-CDW develops longer range correlations with increasing field, but does not increase in amplitude or occupy a significantly larger fraction of the sample volume. The integrated intensity of AF-CDW order does not significantly increase over a range $0\sim16$ T, which leads us to the similar conclusion for the development of AF-CDW. \\

\textbf{Discussion}\\
%\textit{Modulation specific competition:}
The zero-field maximum at $T_\textrm{c}$ of the $\ell=1/2$ CDW diffraction intensity has previously been the subject of different interpretations~\cite{LiuNatComm2016,Chang12a,Hayward14,Nie15}. Our view is that the intensity maximum is the result of phase competition between $\ell=1/2$ CDW order and superconductivity. 
As diffraction intensities are proportional to the square of the CDW order parameter \cite{Jaramillo09a,Huecker14}, temperature (or field) dependence reflects the evolution of the CDW order.
Upon cooling below $T_\textrm{c}$, the superconducting order gradually grows, resulting in a partial suppression of 
the CDW. The balance of this competition can be tipped in favor of the CDW order by application of an external $c$-axis  magnetic field that 
acts to suppress the superconducting order parameter.
With increasing field-strength, the $T_\textrm{max}$ temperature scale shifts to lower values (Fig.~\ref{fig:maps}c) delineating a line in the 
magnetic field phase diagram that provides a firm lower bound for the border between superconductivity and normal state. 
%As such, it can be viewed as the upper critical field line $B_{c2}^{\ell=1/2}$. 
An approximate extrapolation to the $T\rightarrow 0$ limit yields an upper critical field consistent with that inferred from most bulk measurements~\cite{Ramshaw12a,Grissonnanche14,Marcenat15,Zhou17}. It should be stressed that there is still an ongoing debate about the exact field scale of $H_\textrm{c2}$~\cite{Yu16}.
A similar interpretation can be applied to the temperature dependence of the $\ell=1$ response.
As shown in Fig.~\ref{fig:maps}c, the $\ell=1$ maximum temperature scale $T_{\textrm{max}}^{\ell=1}$ is different from $T_\textrm{max}$ for $\ell=1/2$.
This implies the existence of two superconducting onset temperatures at a given applied field or for a fixed temperature two upper critical fields $B_\textrm{max}^{\ell=1/2}$ and $B_\textrm{max}^{\ell=1}$.
The different competing temperature and magnetic field scales observed at $\ell=1/2$ and $\ell=1$ lend support to our starting hypothesis, namely that there are two  different CDW order constituents. In what follows, we discuss the relationship between the CDW and the Fermi surface reconstruction. 

It is interesting to consider the implications  of our correlation length measurements for the electronic band structure~\cite{Gannot2019,BriffaPRB2016}. Although the unfolded structure has not been been resolved, the Fermi surface reconstruction (FSR) revealed by quantum oscillation~\cite{DoironLeyraud07a,Sebastian15,Yao11} and transport~\cite{LeBoeuf07a,LeBoeuf11a,Chang10a} experiments is undoubtedly linked to the appearance of CDW order. It has, however, been debated whether the reconstruction is triggered by the F-CDW or AF-CDW or whether the CDW order is triggered by FSR. The fact that ARPES does not see any FSR~\cite{Hossain08a} has been used as an argument for a field-induced electronic transformation, rather than the reconstruction being already present at zero field. However, the sign change of the Hall effect persists down to as low as 5 T~\cite{Cyr-Choiniere17}, casting doubt on whether the high-field F-CDW order is responsible for the reconstruction.
Recent detailed considerations of CDW correlations concluded that the $\ell=1$ CDW order has about the right in-plane correlation length to explain the quantum oscillation experiments whereas the AF-CDW correlations are too short~\cite{Gannot2019}. We note that at $B=16$ T the in-plane correlation lengths of the F-CDW and AF-CDW correlations are 300 and 100~\AA, respectively. Another signature of FSR is the Hall coefficient which changes sign from positive to negative on cooling~\cite{LeBoeuf07a,LeBoeuf11a,Chang10a,Laliberte11a}.
%Contrasting  arguments have emphasized the Hall coefficient, that evidence the FSR through a sign change from positive to negative values upon cooling. 
It has been demonstrated that for YBCO $p\approx0.12$,  the characteristic sign-changing temperature %scale 
$T_B=66K \simeq T_\textrm{c}$ is essentially independent of magnetic field~\cite{LeBoeuf07a}.
The sign change in the Hall effect persists down to much lower fields, $B=5$\;T, than the quantum oscillations mentioned above. Thus it is interesting to consider whether the F-CDW correlations might be involved in the FSR at lower fields. We have previously demonstrated~\cite{Chang16} that for high magnetic fields, $B=16.5$~T, $\xi^{\ell=1}_b  > \xi^{\ell=1/2}_b$ for $T \approx 66\;K \approx T_B$. We can see from Fig.~\ref{fig:maps}d that this condition persists at $T_{\textrm{max}}$ as $B \rightarrow 0$ and $T_{\textrm{max}} \rightarrow T_B$. In fact, we find that $\xi^{\ell=1}_b$ is longer than $\xi^{\ell=1/2}_b$ down to 5\;T where $T_{\textrm{max}} \approx 55$\;K. Thus, F-CDW correlations with $\xi^{\ell=1}_b \geq 100$~\AA~$ \approx \xi^{\ell=1/2}_b$ exist for temperatures $T \lesssim T_B$ and to lowest measured magnetic fields. Considering the $(B,T)$ phase space, both the F-CDW and AF-CDW correlations could be associated with the Fermi surface reconstruction inferred from Hall effect experiments. However, in both cases the correlation lengths are shorter than the expected threshold for quantum oscillations to be observed. Therefore, either the two-dimensionality of the CuO$_2$ enables the electronic structure to fold also for intermediate correlation lengths or the Fermi surface reconstruction has an entirely different causality and charge order is consequence rather than a trigger.
A central part of this work, is the experimental probing of $\ell=1$ in-plane correlations as a function of magnetic field and temperature. With this, the interaction between $\ell=1$ CDW order and superconductivity can be discussed. 
%In comparison to the extensively studied $\ell=1/2$ order~\cite{Chang12a,Blackburn13a,Huecker14,Ghiringhelli12a,BlancoCanosa13a,Blanco-Canosa14}, 
The $\ell=1$ order displays different competing field and temperature scales, 
compared to the extensively studied $\ell=1/2$ order~\cite{Chang12a,Blackburn13a,Huecker14,Ghiringhelli12a,BlancoCanosa13a,Blanco-Canosa14}. The $T_\textrm{max}^{\ell=1}$ and  $T_\textrm{max}^{\ell=1/2}$ temperature scales indicate the point where superconductivity is strong enough to partially suppress the respective CDW orders. In presence of large enough magnetic fields $B \geq 14$~T, we find that $T_\textrm{max}^{\ell=1}<T_\textrm{max}^{\ell=1/2}$ (Fig.~\ref{fig4}e). 
%This is especially clear for $B \geq 14$~T . 
Extrapolating $T_\textrm{max}^{\ell=1}\rightarrow 0$, a field scale significantly smaller than $B_\textrm{c2}$, inferred from transport measurements~\cite{Grissonnanche14,Ramshaw12a,Chang10a}, is reached. In contrast, $T_\textrm{max}^{\ell=1/2}$ may track $B_\textrm{c2}$ at high fields.
%It is also worth noticing that $T_{max}^{\ell=1}<T_{max}^{\ell=0.5}$ in the $H\rightarrow0$ limit. Therefore, whereas $T_{max}^{\ell=0.5}\rightarrow0$ for $H\rightarrow0$, 
%$T_{max}^{\ell=1}<T_c$.
This suggests that superconductivity occuring in regions with $\ell=1$ CDW order is weaker than in $\ell=1/2$ regions. This is especially the case when the $\ell=1$ correlations develop coherence along the $c$-axis.
The emergence of ferro-coupled long-range $\ell=1$ CDW order, therefore, induces a fragile state of superconductivity. In presence of quenched disorder, such a fragile state of superconductivity has, in fact, been predicted theoretically~\cite{YuPRB2019}. 
%It consists of coupled regions in which superconductivity is partially suppressed. 
Disorder may therefore play an important role for the organization of the observed phase diagram~\cite{YuPRB2019,CaplanPRL2017} and the interaction between CDW order and superconductivity. The exact role of different types of disorder (oxygen-chain vacancies, impurities and vortices) on the AF-CDW order and the fragile superconducting state should be clarified in future experiments. 
The regime with competition induced weakening of superconductivity has been the setting for predictions of pair-density-wave formations~\cite{Agterberg2019,AgterbergNatPhys2008,PepinPRB2014,DaiPRB18,WangPRB18,BergPRL2007}. The existence of such a state should manifest itself through a $\mathbf{Q}_\textrm{PDW}=\mathbf{Q}_\textrm{CDW}/2$ modulation. Although STM studies have reported evidence for PDW order in the halos of vortices~\cite{2018arXiv180204673E}, there are no x-ray diffraction results supporting the existence of a PDW order in YBCO. 

Finally, we comment on the  $S$-shape of $B_\textrm{DOS}$ versus temperature reported by density-of-states sensitive probes (specific heat, thermal conductivity, and NMR)~\cite{KacmarckPRL2018}. For $B>B_\textrm{DOS}$, the density-of-states is magnetic-field independent and the $S$-shape, refers to an effect where $B_\textrm{DOS}$, in addition to standard Bardeen-Cooper-Schrieffer (BCS) behaviour, displays an sudden increase in the $T\rightarrow$~0 limit. Our results do not exclude the possibility that $B_\textrm{max}^{\ell=1}$ would increase steeply in the $T\rightarrow0$ limit. In fact, it is possible that the $\ell=1/2$ and $\ell=1$ superconducting flavors couple in the  $T\rightarrow0$ limit to form a uniform condensate. If so, $B_\textrm{max}^{\ell=1}$ and $B_\textrm{max}^{\ell=1/2}$ should merge for $T=0$. This would lead $B_\textrm{max}^{\ell=1}$ to rise in the low-temperature limit and give it an $S$-shaped dependence reminiscent of that reported by thermodynamic and spin susceptibility probes of $B_\textrm{c2}$~\cite{KacmarckPRL2018}.\\

\textbf{Methods}

\textit{Experimental details:} A high quality single crystal of \ybcosss\ (${T}_{c}=67$ K) with ortho-VIII oxygen ordering was grown and detwinned as described in Ref.~\cite{Liang12a}. Hard x-ray (100 keV) diffraction  experiments were carried out, with a horizontal 17~Tesla magnet~\cite{Holmes12}, at PETRA III's P07 triple-axis diffractometer at DESY (Hamburg, Germany). 
%using a horizontal 17~Tesla magnet~\cite{Holmes12}. 
The YBCO sample was mounted with the magnetic field along the $c$-axis direction giving access to the 
%$\bold{Q}=(0,k,\ell)$ or  
$\bold{Q}=(h,0,\ell)$ scattering plane. The setup is identical to that described in Ref.~\cite{Chang16}, with the exception of an improved sample cooling power allowing a base temperature of $T\approx3$~K to be reached. Scattering vectors are specified in orthorhombic notation with reciprocal lattice units (r.l.u) $(2\pi/a,2\pi/b,2\pi/c)$ where $a \simeq 3.82$~$\rm \AA$, $b \simeq 3.87$~$\rm \AA$, and $c \simeq 11.7$~$\rm \AA$. \\[2mm]

\textbf{Data availability}

All experimental data are available upon request to the corresponding authors. The source data underlying Figs.~2-6, and Supplementary Fig.~1-4 are provided as a Source Data file.\\[2mm]

\textbf{Acknowledgements}

We acknowledge informative discussions with Steve Kivelson. This work was supported by the EPSRC (grant numbers EP/R011141/ \& EP/J016977/1), the
U.S. Department of Energy (DOE), under Contract No. DE-AC02-98CH10886, the Wolfson Foundation, the Royal Society, the Office of Basic Energy Sciences, Division of Materials Science and Engineering, the Danish Agency for Science, Technology and Innovation under DANSCATT, the Leverhulme Trust, and the Swiss National Science Foundation. Parts of this research were carried out at beamline P07 at DESY, a member of Helmholtz Association HGF. \\[2mm]

\textbf{Author contributions}

RL, DAB, and WNH grew and characterised the YBa$_2$Cu$_3$O$_{6.67}$ single crystal. EB, OI, ATH, NBC, MH, EMF, OG, UR, MvZ, SMH and JC executed the x-ray diffraction experiments. JC, OI, SMH, and JC analyzed the data. All authors contributed to the manuscript. J. Choi and O. Ivashko equally contributed to this work. \\

\textbf{Competing interests} 

The authors declare no competing interests.\\

%\bibliography{PRBbib}
% \bibliographystyle{apsrev}
\bibliography{YBCO-NatComm-Final}

\newcommand{\beginsupplement}{
        \setcounter{table}{0}
        \renewcommand{\thetable}{S\arabic{table}}
        \setcounter{figure}{0}
        \renewcommand{\figurename}{\textbf{Supplementary Figure}}}

\beginsupplement
\clearpage
\onecolumngrid
\section{Supplementary Information}

\begin{figure}[htb]
 	\begin{center}
 		\includegraphics[width=0.81\textwidth]{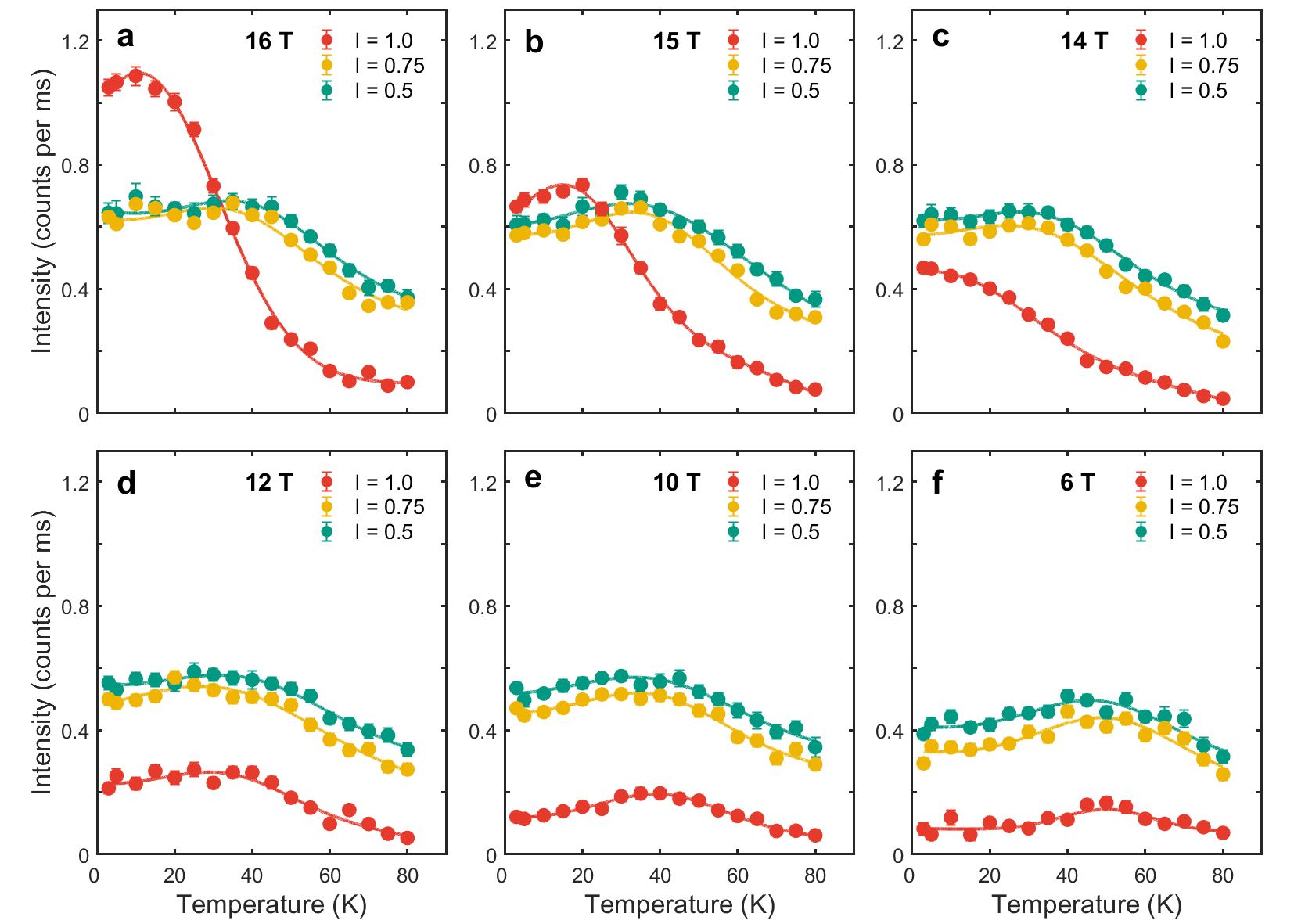}
 	\end{center}
 	\caption{\textbf{Temperature dependence of diffraction intensity at different $\ell$ positions.} Diffraction intensity measured at $\ell$ = 0.5 (green), 0.75 (yellow), and 1 (red) are plotted for comparison as a function of temperature at different magnetic fields as indicated: (a) 16 T (b) 15 T (c) 14 T (d) 12 T (e) 10 T and (f) 6 T. Here, $\ell=0.5$ and $\ell=1$ data are taken from \ref{fig4}b and \ref{fig4}c, respectively. For $\ell=0.75$, the diffracted intensity is integrated and averaged over a range of $0.65<\ell<0.85$ from $\ell$-scans after subtraction of linear backgrounds. Source data are provided as a Source Data file.} \label{fig:S2}
\end{figure}

\begin{figure}[htb]
 	\begin{center}
 		\includegraphics[width=0.9\textwidth]{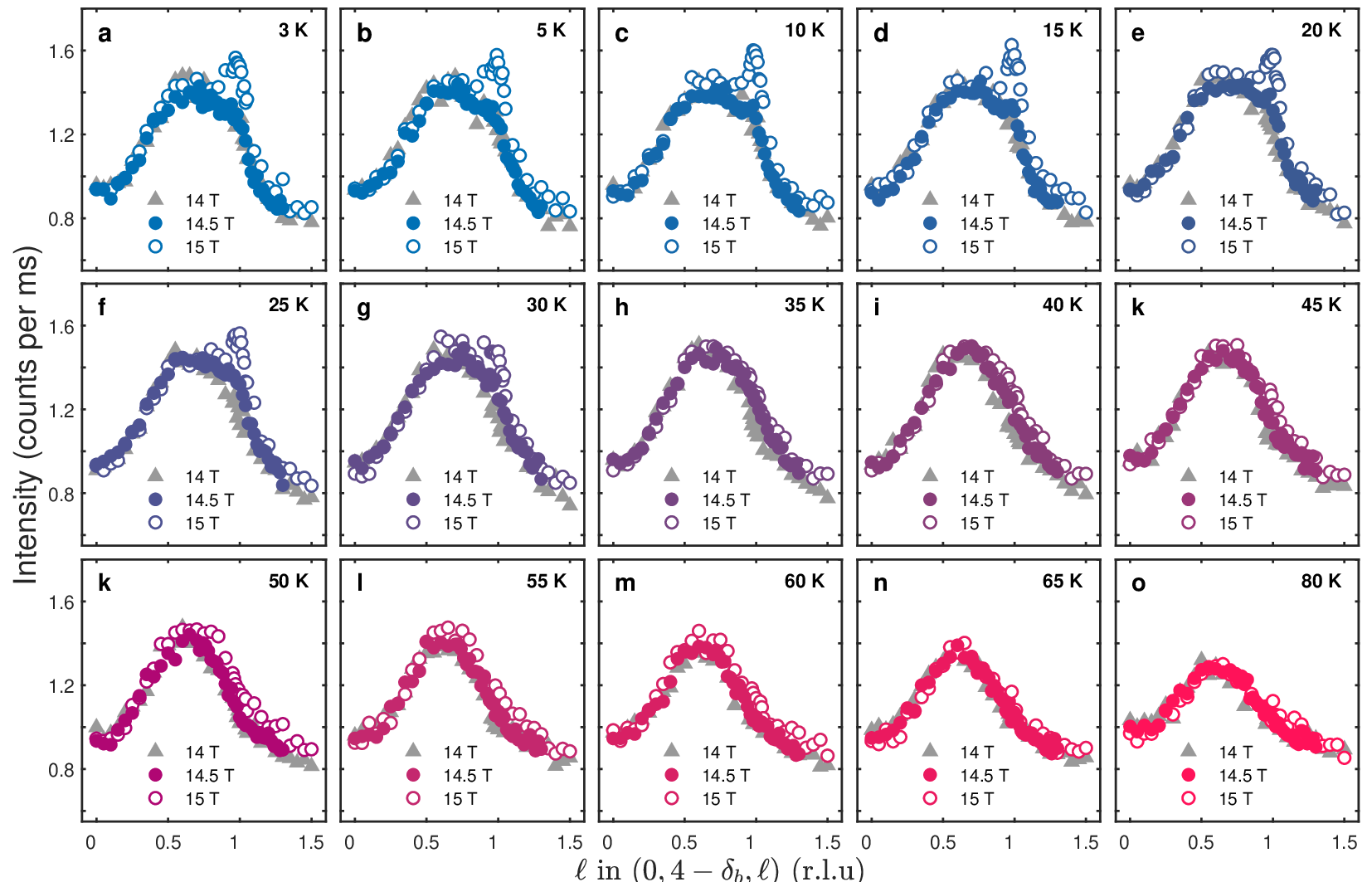}
 	\end{center}
 	\caption{\textbf{Out-of-plane scans ($\ell$-scans) through ($0,4-\delta_b$,1).} \textbf{(a-o)} Raw $\ell$-dependent intensity profiles are plotted at various temperatures as indicated. The data measured at $B=14$ T (grey triangles), 14.5 T (solid circles) and 15 T (open circles) are superimposed at each temperature. Source data are provided as a Source Data file.}\label{fig:S1}
\end{figure}

\begin{figure}[htb]
 	\begin{center}
 		\includegraphics[width=0.9\textwidth]{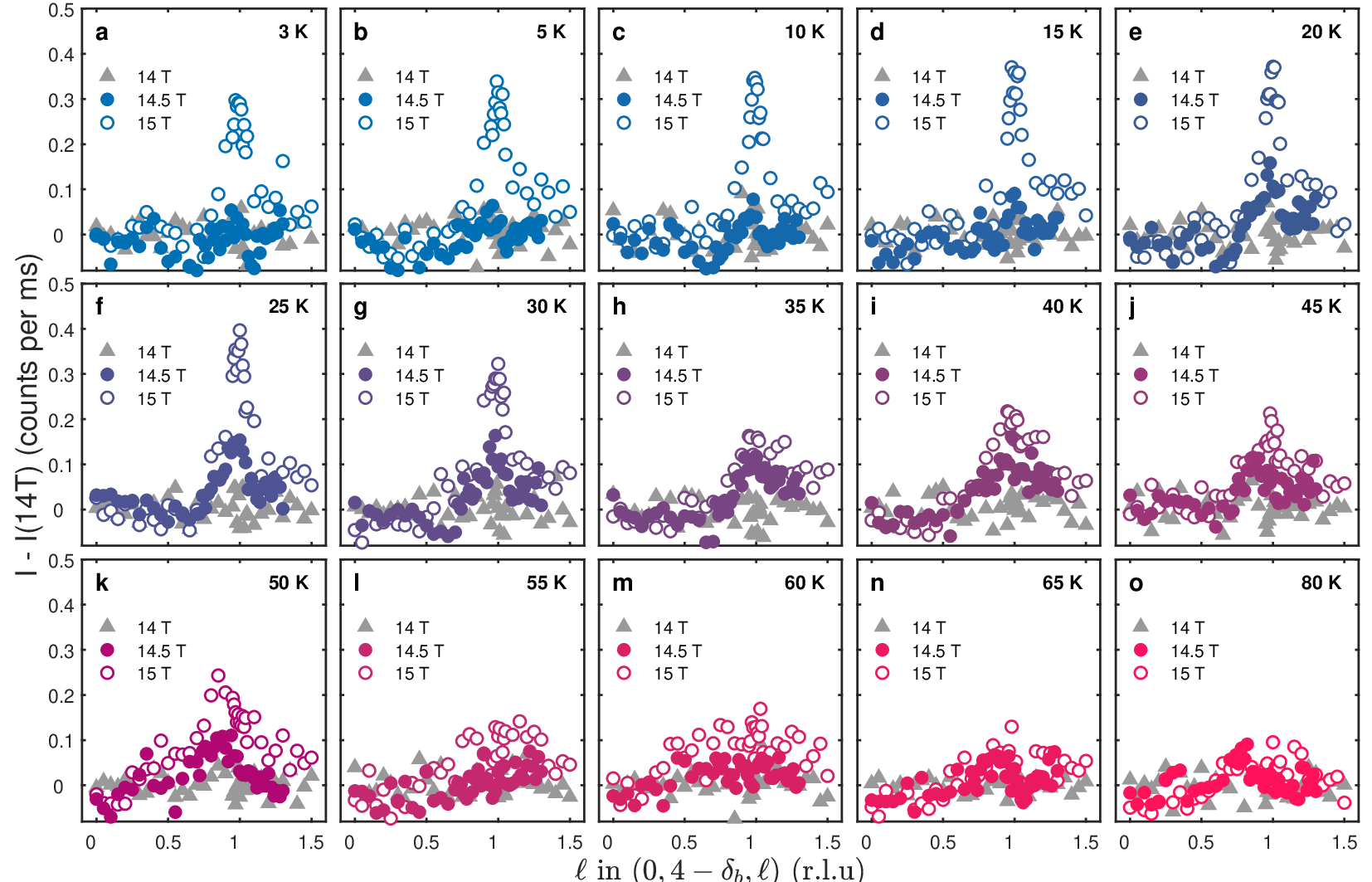}
 	\end{center}
 	\caption{\textbf{Background subtracted $\ell$-scans through ($0,4-\delta_b$,1).} \textbf{(a-o)} background subtracted $\ell$-dependent intensity profiles are plotted at various temperatures as indicated. Gaussian fits of 14-T data are used as the backgrounds at each temperature. Source data are provided as a Source Data file.}\label{fig:S1}
\end{figure}

\begin{figure}[htb]
 	\begin{center}
 		\includegraphics[width=0.85\textwidth]{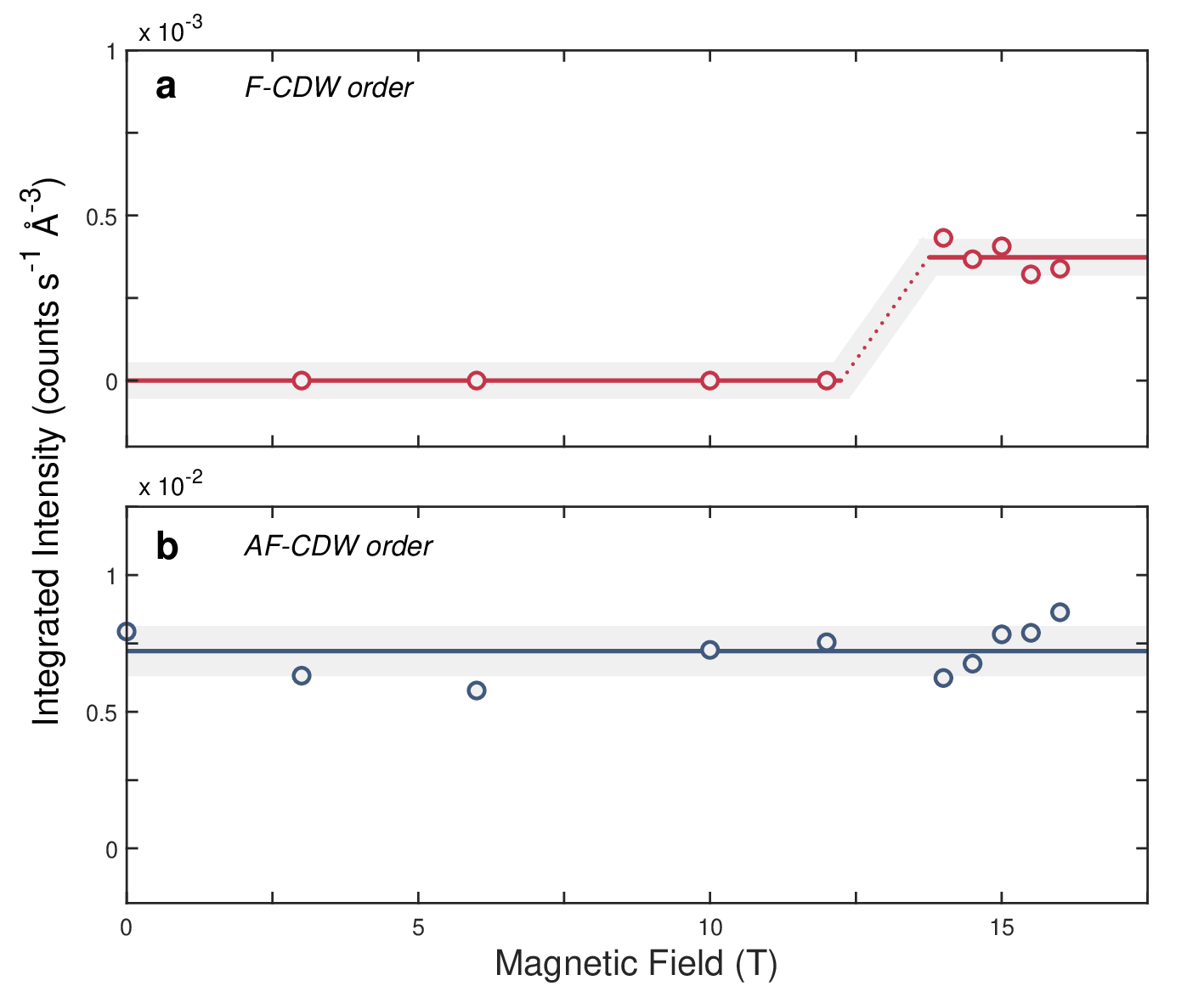}
 	\end{center}
 	\caption{\textbf{Integrated intensity of charge density wave order reflections.} Intensity of \textbf{(a)} ferro and \textbf{(b)} antiferro-coupled charge density wave (CDW) reflections integrated in the three-dimensional reciprocal space, calculated by $I_{\text{CDW}} \times \xi_b^{-2} \times \xi_c^{-1}$, is plotted against a magnetic field. Here, $I_{\text{CDW}}$ is the peak intensity at $(0, 4-\delta, 1/2)$ (AF-CDW) and $(0, 4-\delta, 1)$ (F-CDW) measured at 3 K. $\xi_b$ and $\xi_c$ are the correlation lengths of CDWs along the crystallographic $b$- and $c$-axis direction at 22 and 3 K, respectively. Due to the instrumental constraint, we have limited information about the CDW correlation lengths in the $a$-axis direction, but they appear to be similar to those along the $b$-axis. We have therefore used $\xi_b^{2}$ as a substitute for $\xi_a\times\xi_b$ to estimate the reciprocal-space integrated intensity. Source data are provided as a Source Data file.}\label{fig:S3}
\end{figure}

\end{document}